\documentclass[10pt,english,aps,prd,preprintnumbers,superscriptaddress,floatfix,showpacs,nofootinbib]{revtex4-1}
\usepackage{graphicx}
\usepackage[dvipsnames]{xcolor}
\usepackage{amsmath}
\usepackage{hyperref}

\usepackage[normalem]{ulem}  
\usepackage{ulem}

\usepackage[utf8]{inputenc}  

\newcommand{\be}{\begin{eqnarray}}
\newcommand{\ee}{\end{eqnarray}}
\newcommand{\non}{\nonumber\\}

\newcommand{\ave}[1]{\left\langle #1 \right\rangle}

\newcommand{\mev}{{\rm \, MeV}}

\newcommand{\cb}{CBWC}

\newcommand{\vk}[1]{ }
\newcommand{\BF}[1]{}

\renewcommand\sout{\bgroup\color{blue} \ULdepth=-.5ex \ULset}

\begin{document}

\title{To bin or not to bin: does binning in multiplicity reliably \\ suppress unwanted volume fluctuations?}

\author{Bengt Friman}
\email[E-Mail:]{b.friman@gsi.de}
\affiliation{GSI Helmholtzzentrum f{\"u}r Schwerionenforschung, 64291 Darmstadt, Germany               
}
\author{ 
Volker Koch
}
\email[E-Mail:]{vkoch@lbl.gov}
\affiliation{
Nuclear Science Division,       
Lawrence Berkeley National Laboratory, 1 Cyclotron Road,
Berkeley, CA 94720
}
\date{\today}

\begin{abstract}
  In this study, we examine the effect of the so-called Centrality Bin Width Correction (CBWC) on the measurement of (net-)proton number cumulants in nucleus-nucleus collisions. We present an analytically tractable model, which includes correlations between multiplicity and proton number similar to those generated by the decay of baryon resonances. Within this model, we analyze the circumstances under which the CBWC method correctly removes the undesired effects of volume or impact parameter fluctuations. Additionally, we explore situations where the method fails and produces misleading results. 
\end{abstract}

\maketitle

\section{Introduction}\label{sec:intro}

Exploring the structure of the phase diagram of QCD and searching for possible phase transitions is
one of the major goals in studies of strongly interacting matter. One of the key observables are fluctuations of globally
conserved charges, most prominently (net) baryon number fluctuations in a subsystem of the fireball created in high-energy nucleus-nucleus collisions. In the grand canonical
ensemble, the cumulants of the
net-baryon number distribution are directly related to the derivatives of the pressure with
respect to the baryon chemical potential. Thus, any structure in the partition function due to a possible
phase transition is reflected in the cumulants. However, in order to extract meaningful information
from the cumulants measured in nucleus-nucleus experiments, several effects need to be understood. For example, one needs to correct for the global conservation laws of, e.g., the baryon number
\cite{Bleicher:2000ek,Bzdak:2012an,Vovchenko:2020tsr,Vovchenko:2020gne,Braun-Munzinger:2020jbk,Braun-Munzinger:2023gsd}. One
also needs to account for the fact that only protons and not all baryons are measured
\cite{Kitazawa:2011wh,Kitazawa:2012at}. Finally, even with the tightest centrality cuts, the size of
the system fluctuates, giving rise to so-called volume fluctuations
\cite{Jeon:2003gk,Skokov:2012ds}. At present, two methods have been proposed to remove or at least
minimize the latter. One recent proposal uses the moments of the multiplicity
distribution \cite{Rustamov:2022sqm,Holzmann:2024wyd}. The other, which has been applied by the STAR
collaboration in all analyses, is the so-called centrality bin width correction (CBWC)
\cite{Luo:2013bmi}. In this paper, we study the latter correction in some detail.

We start with some general remarks on the effect of volume or participant fluctuations in nucleus-nucleus collisions. 
In heavy-ion experiments, the centrality of an event is typically defined by the number of charged particles, $n_{ch}$,
produced in the collision. The higher $n_{ch}$ is, the more central is the collision, i.e., the smaller is the (average) impact
parameter. The data are then analyzed for various centrality classes, which are
defined by an interval in multiplicity, $M_{1}\leq M \leq M_{2}$. For example, for the centrality class of the $5\%$
most central collisions, $\int_{0}^{M_{1}} dM P(M) = 0.95 $ and
$\int_{M_{1}}^{M_{2}} dM P(M) = 0.05 $, where $P(M)$ is the
distribution of multiplicities in minimum bias events. Of course, in this case, $M_{2}=M_{max}$ is the maximum
observed multiplicity. Since the charged particle multiplicity in heavy ion reactions is typically
on the order of a few hundred, a centrality class involves a large range of multiplicity values,
$M_{2}-M_{1}\gg 1$. Having a whole range of multiplicities contributing to a centrality class is
actually a welcome feature. Consider a thermal system at fixed ``volume'' where the number of
particles follows a Poisson distribution. If the mean multiplicity is $M=100$ then the system will
contribute significantly at least to events in the multiplicity range $[ M - 10, M + 10]$. Thus, cutting too tight on the multiplicity will
bias the system by suppressing the fluctuations one is interested in. By the same token,
systems with different volumes will contribute to a given centrality
class. \footnote{We note that recently a novel method based on the Edgeworth expansion has been proposed \cite{Wang:2025fve}. It is claimed that this method is able to extract cumulants without a specific centrality selection.} 

The inability to fix the volume in heavy-ion collisions
is the origin of volume fluctuations. The challenge is then to eliminate these from the fluctuation
measurements while simultaneously preserving the physics of interest. In the
following, we will explore to which extent the centrality bin width correction (CBWC) can achieve this. We note that volume fluctuations induce correlations between the multiplicity and
the proton number, since both typically scale with the volume. Consequently, correlations generated by mundane event-by-event variations in collision geometry compete with and may even overshadow
dynamical correlations, such as those due to the decay of baryon resonances into protons and pions.
The primary objective of this paper is to assess the reliability of the CBWC method by employing an analytically tractable model that simulates the correlation generated by baryon resonance decays. While the model has some essential features that allow us to identify and illustrate potential shortcomings of this scheme, it is not sufficiently realistic to allow a direct confrontation with experimental data.

This paper is organized as follows: In the subsequent section, we define the CBWC method and also introduce
the notation. We then describe the model which we use to study the CBWC corrections and present the results, first for the case of proton cumulants and then extend the discussion to net-proton cumulants. We conclude with a discussion and summary of the main findings. With the aim to improve the readability of the main body of the paper, we have relegated  mathematical details to several Appendices.

\section{Definitions and Notation}\label{sec:definitions}
We start by specifying what the Centrality Bin Width Correction (CBWC) entails. 
Consider a given centrality class defined by a multiplicity interval $[ M_{1},M_{2}]$.  Next split
this interval into $n$ bins, $\delta M[i]$, with $1\leq i\leq n$ and consider the events with
multiplicity in a given bin $\delta M[i]$. If the number of events contributing to this bin is
$r_{i}$ then 
\begin{align}
p_{i}=\frac{r_{i}}{\sum_{i=1}^{n} r_{i}}
  \label{eq:bin_prob }
\end{align}
is the fraction of events in the centrality class belonging to bin $\delta M[i]$. Clearly
\begin{align}
\sum_{1}^{n} p_{i} = 1\, ,
  \label{eq:bin_prob_norm}
\end{align}
implying that $p_{i}$ can be considered as the probability that a given event of the centrality class
belongs to bin  $\delta M[i]$. For the events in a bin $\delta M[i]$ one can determine
the moments  $\ave{N^k}_i$ and cumulants $\kappa_{k,i}$  of the particle number distribution of
interest. Given these moments and cumulants together
with the fractions, $p_{i}$  the CBWC method determines the ``centrality bin width corrected'' cumulants or
moments as
\begin{align}\label{eq:CBWC_define}
  \kappa_{n}^{\cb}&=\sum_{i=1}^{n} p_{i}\,\kappa_{n,i}\, , \\
  \ave{N^k}^{\cb} &= \sum_{i=1}^{n} p_{i} \ave{N^k}_{i} \, .
\end{align}

More generally, it will be convenient for our discussion to consider the joint probability
$P(N,M)$ to find $N$ particles for a given multiplicity $M$ for the centrality class of interest. With
$\sum_{N,M}P(N,M)=1$ the multiplicity distribution, $P(M)$, in the centrality class is then
\begin{align}
P(M) = \sum_{N}P(N,M)\,.
  \label{}
\end{align}
Therefore, the probability, $p_{i}$, for an event to belong to bin $i$, Eq.~\eqref{eq:bin_prob }, is
\begin{align}
p_{i}=\sum_{M\in \delta{M[i]} } P(M)\, .
  \label{eq:bin_prob}
\end{align}
The particle number distribution, $P(N)$, for the entire event class is
\begin{align}
P(N) =\sum_{M}P(N,M) \, ,
   \label{eq:full-dist}
\end{align}
while the (normalized) probability distribution for particles in bin $i$ is given by
\begin{align}
    P_i(N)=\frac{\sum_{M\in \delta M[i]} P(N,M)}{\sum_{N}\sum_{M\in \delta{M[i]} } P(N,M) } =
  \frac{1}{p_{i}} \sum_{M\in \delta M[i]} P(N,M)\,.
\label{eq:part_dist_bin}
\end{align}

It is useful to introduce the  moment and cumulant generating
functions. For the entire
multiplicity class, they are denoted by $H(t)$ and $G(t)=\ln\left[ H(t) \right]$, respectively. Then the moments, $\ave{N^{k}}$ and cumulants
$\kappa_{k}$ are given by
\begin{align}
\ave{N^{k}} = \left.\frac{\partial^{k}}{\partial t^{k}}H(t) \right|_{t=0}\, ,\;\;\;\ \kappa_{k}=\left.\frac{\partial^{k}}{\partial t^{k}}G(t) \right|_{t=0}\, .
  \label{eq:gen_func_full}
\end{align}
The generating functions for the events in a given bin $i$, which we denote by $h_{i}(t)$ and $g_{i}(t)=\ln\left[ h_{i}(t) \right]$, yield the moments and
cumulants in that bin
\begin{align}
\ave{N^{k}}_{i} = \left.\frac{\partial^{k}}{\partial t^{k}}h_i(t) \right|_{t=0}\, ,\;\;\;\ \kappa_{k,i}=\left.\frac{\partial^{k}}{\partial t^{k}}g_i(t) \right|_{t=0}\, .
  \label{eq:moments_bin}
\end{align}
With the above expressions for the particle distribution in the entire multiplicity class and in a
bin $i$, the corresponding moment generating functions are 
\begin{align}
  H(t) = \sum_{N} P(N) e^{N \, t}\, ,
  \label{eq:mom_gen_full}
\end{align}
and
\begin{align}
h_{i}(t) = \sum_{N} P_{i}(N) e^{N \, t} \, .
  \label{eq:mom_gen_bin}
\end{align}

In order to calculate net proton cumulants, we need a probability distribution for the number of protons and antiprotons, $P(N,\bar{N})$. 
It is convenient to first compute the mixed cumulants $C^{(n,m)}$ of this distribution and then obtain the net proton cumulants using the general relation \cite{Friman:2022wuc} 
\begin{equation}\label{eq:net-cumulant}
    \kappa^{net}_n=\sum_{i=0}^n \binom{n}{i}\,(-1)^{n-i}\,C^{(i,n-i)}\, .
\end{equation}
Accordingly, we first define generalized generating functions 
\begin{align}\label{eq:mixed-gen-func}
    H(t,s)=\sum_{N=0}^{\infty}\sum_{\bar{N}=0}^{\infty}P(N,\bar{N})\,e^{N\,t+\bar{N}\,s}\, ,\qquad
    G(t,s)=\ln\big[H(t,s)\big]\, ,
\end{align}
and then compute the mixed moments and cumulants using~\footnote{The net proton cumulants can be computed directly from the cumulant generating function  \cite{Bzdak:2012an,Braun-Munzinger:2020jbk,Friman:2022wuc} using $\kappa_n=\frac{d^n}{d\,t^n}\,G(t,-t)|_{t=0}$. However, the slight detour we follow here allow us to explore also the consequences of CBWC on the mixed cumulants with little extra effort.}
\begin{align}
    \ave{N^n\,\bar{N}^m}&=\frac{\partial^n}{\partial t^n}\,\frac{\partial^m}{\partial s^m}\,\left.H(t,s)\right|_{t=0,s=0}\, ,\\
    C^{(n,m)}&=\frac{\partial^n}{\partial t^n}\,\frac{\partial^m}{\partial s^m}\,\left.G(t,s)\right|_{t=0,s=0}\, .\label{eq:mixed-cumulants}
\end{align} 

\section{General observations}
Before we discuss the detailed calculation of the CBWC method, let us make some general
observations. First, we re-write the moment generating function $H(t)$
of the full centrality class, Eq. \eqref{eq:mom_gen_full}, as
\begin{align}
  H(t) &= \sum_{N} P(N) e^{N \, t} = \sum_{M} \sum_{N} P(N,M) e^{N \, t} =
  \sum_{i} \sum_{M \in \delta{M[i]}} \sum_{N} P(N,M) e^{N \, t}\\&=  \sum_{i}  \sum_{N} \sum_{M \in \delta{M[i]}} P(N,M) e^{N \, t}\, .\nonumber
  \label{}
\end{align}
Using Eqs. \eqref{eq:part_dist_bin} and \eqref{eq:mom_gen_bin} we then find
\begin{align}
  H(t) = \sum_{i} p_{i} h_{i}(t) \,,
  \label{}
\end{align}
i.e., the moment generating function $H(t)$ is the sum of those of the bins, $h_{i}(t)$, weighted by the corresponding probabilities, $p_i$.
It follows that the moments for the entire centrality class are identical to the CBWC corrected ones,
\begin{align}
\ave{N^{k}} = \left.\frac{\partial^{k}}{\partial t^{k}}H(t) \right|_{t=0} = \sum_{i} p_{i}
  \left.\frac{\partial^{k}}{\partial t^{k}}h(t) \right|_{t=0} = \sum_{i} p_{i }\ave{N^{k}}_{i} = \ave{N^{k}}^{\cb}\, .
  \label{eq:moments_same}
\end{align}
Thus, the CBWC procedure does
not modify the moments of the distribution. However, this is different for the cumulants, as follows directly from the definition of the cumulant generating function. In this case we have
\begin{align}
G(t) = \ln\left[ H(t)  \right] = \ln\left[  \sum_{i} p_{i} h_{i}(t) \right] \ne \sum_{i} 
  p_{i}\ln\left[ h_{i}(t)  \right] = \sum_{i}   p_{i} \, g_{i}(t)\, .
  \label{}
\end{align}

Consequently, in general, the cumulants obtained with the CBWC procedure are different from those for the
entire event class, i.e., 
\begin{align}
\kappa_{k}\ne \kappa_{k}^{\cb}\, .
  \label{}
\end{align}
As an example, consider the second order cumulants, $\kappa_{2}$ and $\kappa_{2}^{\cb}$.
Expressed in terms of moments, they are given by
\begin{align}\label{eq:variance-full}
  \kappa_{2} &= \ave{N^{2}}-\ave{N}^{2}\, , \\
    \kappa_2^{CBWC}&=\langle N^2\rangle^{\cb}-(\langle N\rangle^2)^{\cb}=\langle N^2\rangle-(\langle
    N\rangle^2)^{\cb}\, ,\label{eq:variance-CBWC}
\end{align}
where in the last step, we used Eq. \eqref{eq:moments_same}.
It follows from Eqs. \eqref{eq:CBWC_define}, \eqref{eq:moments_same}, and 
\begin{equation}\label{eq:kappa-2-i}
    \kappa_{2,i}=\langle N^2\rangle_i-\langle N \rangle_i^2\, ,
\end{equation} 
that the final terms in \eqref{eq:variance-full} and \eqref{eq:variance-CBWC} are given by
\begin{align}
  \ave{N}^2=\left(\sum_i p_i\,\ave{N}_i \right)^2\, , \non
  (\ave{N}^2)^{\cb}=\sum_{i} p_i\, \ave{N}_{i}^2\, .
\end{align}
Thus, the difference between the full and CBWC second order cumulants 
\begin{align}
  \Delta_2 &=  \kappa_2 - \kappa_{2}^{CBWC}= \sum_i p_i\, \ave{N}_{i}^2-\left(\sum_i p_i\,\langle N\rangle_i\right)^2
  \label{}
\end{align}
is the variance of the distribution of $\ave{N}_{i}$ over the bins $i$, and is therefore positive
semidefinite. In other words, the CBWC corrected second order cumulant is smaller than or equal to that of
the uncorrected one,   
\begin{equation}\label{eq:variance-inequality}
    \kappa_2-\kappa_2^{CBWC}=\Delta_2\geq 0\,.
\end{equation}
If and only if the means for bins, $\langle N\rangle_i$, are all equal, i.e., independent of $i$, the variance
$\Delta_2$ vanishes and the cumulants (\ref{eq:variance-full}) and (\ref{eq:variance-CBWC}) are identical. \footnote{Analogous conditions apply also for higher-order cumulants.} The inequality (\ref{eq:variance-inequality}) illustrates the fact that in general, fluctuations are suppressed by the CBWC procedure. In order to explore to what extent the CBWC scheme removes the unwanted volume fluctuations and whether it suppresses the physics one is interested in, we perform a more detailed analysis in the following sections.

\section{Analysis of the CBWC procedure}\label{sec:Analysis-CBWC}

After these more general observations, we now discuss the CBWC procedure in detail within an
analytically tractable model. Such a model requires two ingredients: (a) a distribution of the
volumes to model the volume fluctuations, and (b) a joint probability $P(N,M; V)$ which depends on
the volume. The joint probability $P(N,M)$ is then obtained by folding the $P(N,M;V)$ with
the distribution of volumes. In terms of the reduced volume $x=V/V_{0}$ with $V_{0}$ the
average or mean volume, the joint probability $P(N,M)$ for a system with volume fluctuations is
given by
\begin{align}
P(N,M) = \int dx\, Q(x)\, P(N,M;x\, V_0)\, ,
  \label{}
\end{align}
where $Q(x)$ represents the distribution of the reduced volume, $x$. 

In the following we will model the volume fluctuations with a Gamma-distribution and the
joint probability $P(N,M; x)$ with a bi-variate Poisson distribution. The Gamma distribution for
the reduced volume, $Q(x)$, is given by (see Appendix \ref{sec:gamma_dist} for details)
\begin{align}
 Q(x) =\frac{k^{k}}{\Gamma(k)}x^{k-1}e^{-xk} ;\qquad \int_{0}^{\infty}dx\,Q(x) =1\, ,
  \label{eq:gamma_Q}
\end{align}
while the bi-variate Poisson distribution is (see Appendix \ref{sec:bi-variate-poiss} for details)
\begin{align}
    P\left[N, M;\nu,\mu,\lambda\right]&=e^{-(\lambda + \nu + \mu)}\frac{\nu^N}{N!}\,\frac{\mu^M}{M!} \non
    &\times\,\sum_{j=0}^{{\rm Min}(N,M)}j!\,{N\choose j}\,{M\choose j}\,\left(\frac{\lambda}{\nu
      \,\mu}\right)^j\, .
      \label{eq:bivariate-poisson}
\end{align}

Multi-variate Poisson distributions are useful for modeling fluctuations of correlated variables. To see how the bi-variate Poisson distribution arises, consider the
numbers $(n,m,k)$, which are distributed according to three independent Poisson distributions with
means $\nu,\mu,\lambda$, respectively. Then defining $N=n+k$, $M=m+k$ and summing over $k$, keeping $N,M$ fixed, yields the distribution $P\left[N, M;\nu,\mu,\lambda\right]$. A physical example of this would be a system with protons, pions and
$\Delta$ resonances,
all Poisson distributed with means $\nu,\mu,\lambda$. Then $N$ would be the number of
protons and $M$ the number of pions in the final state, after
the decay of the $\Delta$ resonances, $\Delta\rightarrow p+\pi$. Furthermore, the
parameter $\lambda$ controls the strength of the dynamical correlation between $N$ and $M$, which in our
example is realized through the decay of $\Delta$ (or other baryon) resonances into protons and
charged pions. Thus, $\lambda$ corresponds to the mean number of $\Delta^{++}$ and $\Delta^{0}$ resonances.\footnote{Actually this correspondence is not one-to-one, since only 1/3 of $\Delta^0$ contribute owing to the branching ratio for  $\Delta^0 \rightarrow p+\pi^-$.} Note, that $M$ does not contain ``protons''. This is actually a
desired feature since in the analysis by the STAR collaboration the multiplicity is given by
all charged particles {\em except} protons.

\begin{figure}[ht]
    \centering
     \includegraphics[width=0.6\textwidth]{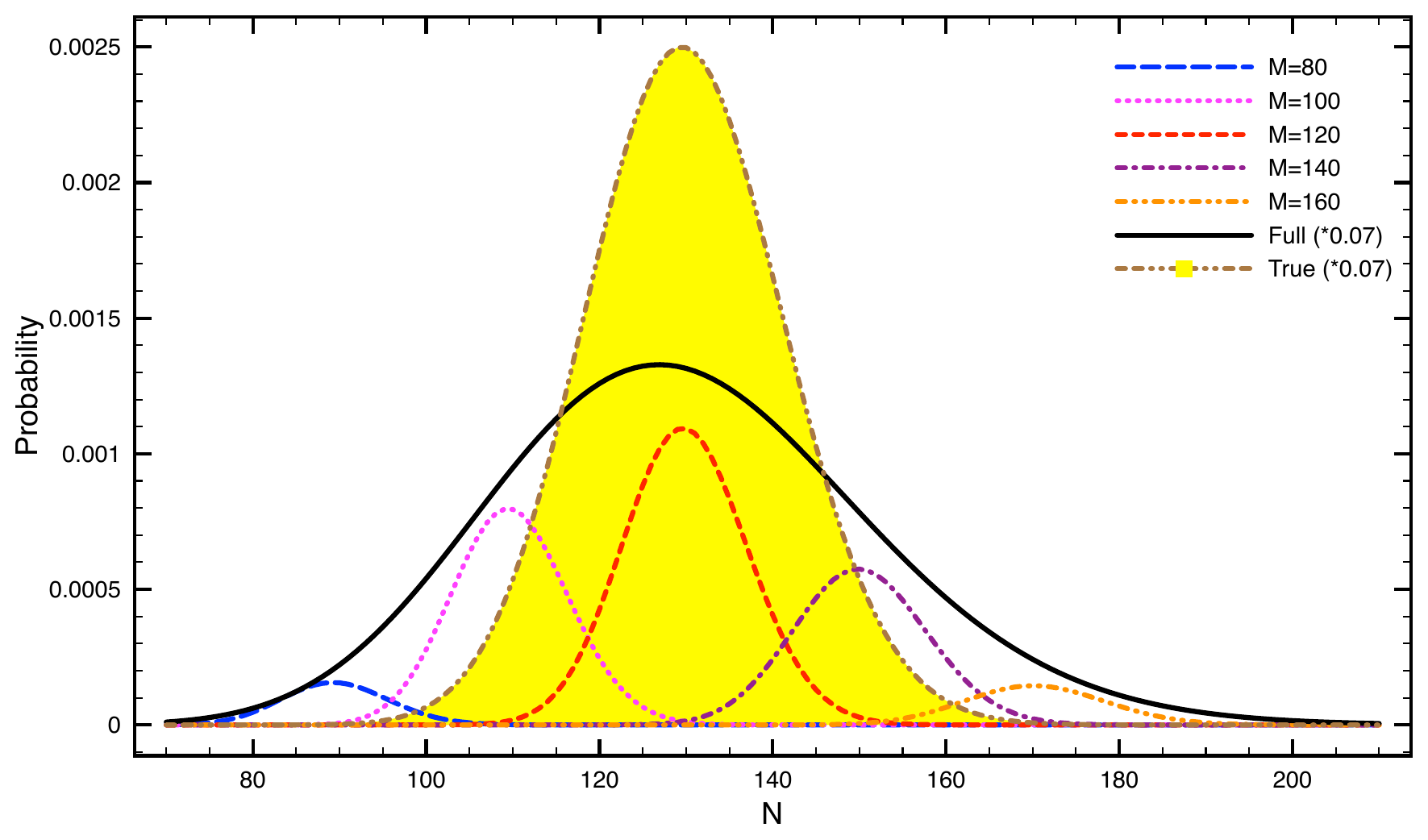}
    \caption{The probability distribution for $N$ \eqref{eq:model_joint_dist} for fixed $M= 80,\,100,\,120,\,140$ and $160$. Also shown are the original distribution, without volume fluctuations \eqref{eq:bivariate-poisson} (filled in yellow) as well as the full distribution, which includes volume fluctuations \eqref{eq:model_joint_dist}, both summed over all $M$. The parameters employed here are $k=50,\,\lambda=100,\,\nu=30,\,\mu=20$.  
    }
    \label{fig:Ndist_fixed_M_and_full}
\end{figure}

The effect of volume fluctuations on the bi-variate Poisson distribution is introduced by
scaling the means with the reduced volume,\footnote{Here we implicitly assume that the means are proportional to the volume. For large systems this is the leading term, but for finite systems there are corrections, the discussion of which is beyond the scope of the present work.} $(\nu,\mu,\lambda) \rightarrow (x\,\nu,x\,\mu,x\,\lambda)$, 
so that our model distribution is
\begin{align}
P(N,M) = \int_0^\infty dx\, Q(x) P\left[N, M; x\,\nu,x\,\mu,x\,\lambda\right]\, .
  \label{eq:model_joint_dist}
\end{align}
In Fig. \ref{fig:Ndist_fixed_M_and_full} we illustrate the dependence of the distribution \eqref{eq:model_joint_dist} on $N$ for several values of $M$. We also show the full distribution, obtained by substituting (\ref{eq:model_joint_dist}) in (\ref{eq:full-dist}), as well as the original (fixed volume) bivariate Poisson distribution \eqref{eq:bivariate-poisson}, summed over all $M$. 

Given the joint distribution $P(N,M)$, Eq.\eqref{eq:model_joint_dist}, we can now proceed to
compute the cumulants as well as their CBWC corrected counterparts. Our starting point is the moment
generating function of the bi-variate Poisson distribution for a given volume fraction $x$ (see
Appendix \ref{sec:bi-variate-poiss})
\begin{align}
  H_{NM}(t,s,x)&= \sum_{N,M}P\left[N, M; x\,\nu,x\,\mu,x\,\lambda\right] e^{N t}e^{M s} \non
    &=\exp\left(-(\lambda+\nu+\mu)\,x\right)\exp\left(e^s(e^t\,\lambda+\mu)\,x+e^t\,\nu\,x\right)\, .
  \label{eq:mom_gen_all}
\end{align}

First, we determine the cumulants for the entire centrality class, i.e., we sum over all
multiplicities. The corresponding moment generating function for a given $x$ is
\begin{align}
  H_{N}(t,x)&= \sum_{N,M}P\left[N, M; x\,\nu,x\,\mu,x\,\lambda\right] e^{N t} = H_{NM}(t,s=0,x) \non
              &= \exp\left(-(\lambda+\nu)\,x\right)\exp\left(e^{t}(\nu+\lambda)\,x \right)\, .
  \label{eq:mom-gen-func-fixed-vol}
\end{align}
Integrating over the distribution of reduced volumes, we obtain the total moment generating
function
\begin{align}
H_{vol}(t) &= \int_0^\infty Q(x) H_{N}(t,x) 
    =\left(\frac{k}{k+(1-e^t)\,(\lambda+\nu)}\right)^k\, .
  \label{eq:moment_gen_full}
\end{align}
The corresponding generating function for the total cumulants is
\begin{equation}
  G_{vol}(t)=\log(H_{vol}(t))=-k\,\log\left(\frac{k+(1-e^t)\,(\lambda+\nu)}{k}\right)\, .
   \label{eq:cumulant_gen_full}
\end{equation}

We note that the generating functions for the case without volume fluctuations are obtained from (\ref{eq:mom-gen-func-fixed-vol}) by setting $V=V_0$, i.e., $x=1$,
\begin{align}
    H_{true}(t)&=H_N(t,x=1)=\exp[(e^t-1)(\nu+\lambda)]\, ,\\
    G_{true}(t)&=\log(H_N(t,x=1))=(e^t-1)(\nu+\lambda)\, .
\end{align}
These are the generating functions of a Poisson distribution with the mean $\nu+\lambda$. Consequently, in our model the true cumulants are 
\begin{equation}\label{eq:cumulants_true}
    \kappa^{true}_n=\nu+\lambda\, ,
\end{equation} 
independent of $n$. It is these cumulants we actually want to extract by eliminating the volume fluctuations.

Using Eq.~\eqref{eq:gen_func_full} and the Fa\'{a} di Bruno's formula~\cite{Riordan:1946,Comtet:1974}, one finds for the total moments and cumulants
\begin{align}\label{eq:full-gamma-moments}
    \ave{N^{n}}&=\sum_{m=1}^n\,\sum_{j=1}^m\,s(m,j)\,\left(\frac{-1}{k}\right)^{m-j}\,B_{n,m}(\lambda+\nu,\lambda+\nu,\dots) \non
    &=\sum_{m=1}^n\,S(n,m)\,\left(\frac{\lambda+\nu}{-k}\right)^m\,\sum_{j=1}^m\,s(m,j)\,(-k)^j\, ,
\end{align}
and
\begin{equation}
  \kappa_n^{vol}=\sum_{j=1}^n\,(j-1)!\,S(n,j)\,k\,\left(\frac{\lambda+\nu}{k}\right)^j\, .
  \label{eq:cumulants_vol}
\end{equation}
Here $s(m,j)$ and $S(n,j)$ are Stirling numbers of the first and second kind, respectively, and for the partial Bell polynomials we have used
\begin{equation}
    B_{n,m}(\alpha,\alpha,\dots,\alpha)=\alpha^m\,S(n,m)\, .
\end{equation}
These cumulants include all effects due to volume fluctuations. 
Explicit forms for the first four cumulants are given below.

In order to apply the CBWC procedure, we need the cumulants of each bin. As detailed in Appendix \ref{sec:cum_calc}, the cumulant generating function for a given bin is  
\begin{align}
    g(t;M_{j})=M_{j}\,\log\left(\frac{e^t\,\lambda+\mu}{\lambda+\mu}\right)-(M_{j}+k)\,\log\left(\frac{k+\lambda+\mu+(1-e^t)\nu}{k+\lambda+\mu}\right)\, .
  \label{eq:cumulant-gen-function}
\end{align} 
Differentiating the cumulant generating function $g(t;M_{j})$ according to Eq.~\eqref{eq:moments_bin}, we
obtain the cumulants for a given bin $M_{j}$ (see Appendix \ref{sec:cum_calc} for details),
\begin{align}
    \kappa_{n,j}
    &=\,\sum_{j=1}^n\,(j-1)!\,S(n,j)\,\left[(k+M_{j})\,\left(\frac{\nu}{k+\lambda+\mu}\right)^j -M_{j}\,\left(\frac{-\lambda}{\lambda+\mu}\right)^j\right]\, .
  \label{eq:cumulant_bin}
\end{align}
The CBWC-corrected cumulants are then obtained by summing the cumulants $\kappa_n[M_{j}]$  over the bins
$M_{j}$, with the proper statistical weight, $p_j$ (see Eq. \eqref{eq:CBWC_define}). 
Equations \eqref{eq:bin_prob_model} and \eqref{eq:h_tilde_final} yield the statistical weights
\begin{align}
 p_{j} =  k(t=0,M_{j}) = \frac{\Gamma(M_{j}+k)}{\Gamma(k)\,\Gamma(M_{j}+1)}
                               \frac{k^k\,(\lambda+\mu)^{M_{j}}}{(k+\lambda+\mu)^{M_{j}+k}}\, .
  \label{eq:p_j_model}
\end{align} 
It follows that $\langle M\rangle=\sum_{j} p_{j} M_{j} = \lambda+\mu$, and consequently that the CBWC cumulants are given by
\begin{align}
    \kappa_n^{CBWC} &= \sum_j p_j\, \kappa_{n,j}
    =\sum_{j=1}^n\,(j-1)!\,S(n,j)\,\left[\frac{\nu^j}{(k+\lambda+\mu)^{j-1}}-\frac{(-\lambda)^j}{(\lambda+\mu)^{j-1}}\right]\, .\label{eq:cumulants_CBWC}
\end{align}

We summarize the results of this section by giving explicit forms for the first four cumulants in the three cases \eqref{eq:cumulants_true}, \eqref{eq:cumulants_vol} and \eqref{eq:cumulants_CBWC}. The first order cumulants, $\kappa_{1}$, are all identical,
\begin{equation}
    \kappa_{1}^{true}=\kappa_{1}^{vol}=\kappa_{1}^{\cb}=\lambda+\nu\, ,
\end{equation}
while for the variance, $\kappa_2$, we have
\begin{align}\label{eq:k2_all}
    \kappa_2^{true}&=\lambda+\nu\, ,&\text{Total variance, no vol. fluct.}\, ,\\\nonumber
    \kappa_2^{vol}&=\lambda+\nu+\frac{(\lambda+\nu)^2}{k}\, ,&\text{Total variance, with vol. fluct.}\, ,\\\nonumber
    \kappa_2^{CBWC}&=\lambda+\nu+\frac{\nu^2}{k+\lambda+\mu}-\frac{\lambda^2}{\lambda+\mu}\,
                     ,&\text{CBWC variance}\, .
\end{align}
For the third cumulants we find,
\begin{align}\label{eq:k3_all}
    \kappa_3^{true}&=\lambda+\nu\, ,\\\nonumber
    \kappa_3^{vol}&=\lambda+\nu+3\,\frac{(\lambda+\nu)^2}{k}+2\,\frac{(\lambda+\nu)^3}{k^2}\, , \non
    \kappa_3^{CBWC}&=\lambda+\nu+3\left(\frac{\nu^2}{(k+\lambda+\mu)}-\frac{\lambda^2}{(\lambda+\mu)}\right)+2\left(\frac{\nu^3}{(k+\lambda+\mu)^2}+\frac{\lambda^3}{(\lambda+\mu)^2}\right)\, ,\nonumber                 
\end{align}
and for the fourth ones,
\begin{align}\label{eq:k4_all}
    \kappa_4^{true}&=\lambda+\nu\, ,\\\nonumber
    \kappa_4^{vol}&=\lambda+\nu+7\,\frac{(\lambda+\nu)^2}{k}+12\,\frac{(\lambda+\nu)^3}{k^2}+6\,\frac{(\lambda+\nu)^4}{k^3} \, ,\non
    \kappa_4^{CBWC}&=\lambda+\nu+7\left(\frac{\nu^2}{(k+\lambda+\mu)}-\frac{\lambda^2}{(\lambda+\mu)}\right)+12\left(\frac{\nu^3}{(k+\lambda+\mu)^2}+\frac{\lambda^3}{(\lambda+\mu)^2}\right)\non
    &+6\,\left(\frac{\nu^4}{(k+\lambda+\mu)^3}-\frac{\lambda^4}{(\lambda+\mu)^3}\right)\, .\nonumber       
\end{align}
We note that the cumulants including volume fluctuations, $\kappa_{n}^{vol}$, are exactly given by the  combination of
the cumulants at fixed volume, $\kappa_{n}^{true}$, and those of the volume distribution,
$\kappa_{n}^{Q}$, Eq.~\eqref{eq:vol-kappa}, as derived in Ref.\cite{Skokov:2012ds}.

Given the proton cumulants, the corresponding proton factorial cumulants can be obtained using the general relation \cite{Friman:2022wuc}
\begin{equation}
    F_n^{a}=\sum_{m=1}^n\,s(n,m)\,\kappa_m^{a}\, ,
\end{equation}
where $a$ is either $true$, $vol$ or $CBWC$. Since the true fluctuations in our model are Poissonian, $F_n^{true}=0$ for $n\geq 2$, while the corresponding $F_n^{vol}$ and $F_n^{CBWC}$ are given by 
\begin{align}
    F_n^{vol}&=(n-1)!\,\frac{(\lambda+\nu)^n}{k^{n-1}}\, ,\\
    F_n^{CBWC}&=(n-1)!\left[\frac{\nu^n}{(k+\lambda+\mu)^{n-1}}-\frac{(-\lambda)^n}{(\lambda+\mu)^{n-1}}\right]\, .
    \label{eq:CBWC-fact-cum}
\end{align}
For more details, see Appendix \ref{sec:factorial-cumulants}.

In the subsequent section, we briefly discuss the CBWC-corrected cumulants for net protons in systems with both protons and antiprotons.

\section{Including antiprotons}
In order to describe the fluctuations at higher energies, we need a probability distribution, where not only the fluctuations of protons but also those of the antiprotons are correlated with the fluctuations of multiplicity. To describe this, we employ a tri-variate Poisson distribution for the number of protons $N$, antiprotons $\bar{N}$, and multiplicity $M$ (see Appendix \ref{sec:appendix_net_prot} for details)
\begin{align}
    P_{TP}\left[N,\bar{N},M\right]
    &=e^{-(\nu+\bar{\nu}+\mu+\lambda+\bar{\lambda})}\,\sum_{l=0}^{{\rm Min}(N,M)} \frac{\nu^{\,N-l}\,}{l!\,(N-l)!}\,\sum_{\bar{l}=0}^{{\rm Min}(\bar{N},M-l)}\frac{\bar{\nu}^{\,\bar{N}-\bar{l}}\,\mu^{\,M-l-\bar{l}}\,\lambda^{\,l}\,\bar{\lambda}^{\,\bar{l}}}{\bar{l}!\,(\bar{N}-\bar{l})!\,(M-l-\bar{l})!}\, .  
\end{align}
The strength of the correlations of the proton and antiproton numbers with multiplicity is governed by the parameters $\lambda$ and $\bar{\lambda}$, respectively.

The moment generating function for this distribution is given by
\begin{align}
    h_{N\bar{N}M}(t,s,u)&=\sum_{N,\bar{N},M}\,e^{N\,t+\bar{N}\,s+M\,u}\,P(N,\bar{N},M)\\
    &=e^{(e^t-1)\nu+(e^s-1)\bar{\nu}+(e^u-1)\mu+(e^{t+u}-1)\lambda+(e^{s+u}-1)\bar{\lambda}}\, .
\end{align} 
The physical picture is analogous to that in the previous case, namely that of a system with (primordial) protons, antiprotons, and pions, which are Poisson distributed with means $\nu, \bar{\nu}$, and $\mu$, respectively, as well as $\Delta$ and anti-$\Delta$ resonances, also Poisson distributed with means $\lambda$ and $\bar{\lambda}$. As before, the resonances decay into pions and baryons or antibaryons, giving rise to correlations between the multiplicity and the proton or antiproton numbers.

The net proton cumulants in this model are obtained by a calculation, which is similar though more involved than that outlined in section \ref{sec:Analysis-CBWC}. As detailed in Appendix \ref{sec:appendix_net_prot}, we first compute the mixed cumulants, Eq.\eqref{eq:mixed_cum_net}, by differentiating the corresponding cumulant generating functions and then obtain the net proton cumulants using the general relation (\ref{eq:net-cumulant}). Here we present only the main results. 

The true mixed cumulants, i.e., those at fixed volume, are in our model given by (see Eq. \eqref{eq:mixed_cum_net_true})
\begin{align} 
    C^{(n,m)}_{true} =(\nu+\lambda)\delta_{m,0}(1-\delta_{n,0})+(\bar{\nu}+\bar{\lambda}
)\delta_{n,0} (1-\delta_{m,0})   \, .
\end{align} 
Since only cumulants of type $C^{(n,0)}_{true}$ and $C^{(0,m)}_{true}$ are non-zero, there is no correlation between the proton and anti-proton fluctuations. Thus, the true model distribution corresponds to two independent Poisson distributions with the means $\nu+\lambda$ and $\bar{\nu}+\bar{\lambda}$, respectively. The fluctuations of the net-proton number are then given by a Skellam distribution, with cumulants  
\begin{equation}\label{eq:net-proton-true}
    \kappa^{true}_n=\nu+\lambda + (-1)^n\,(\bar{\nu}+\bar{\lambda})\, .
\end{equation}

The mixed cumulants, including volume fluctuations, described by the Gamma distribution (\ref{eq:gamma_Q}), are given by (\ref{eq:net-proton-fluct-vol}),
\begin{eqnarray}
    C^{(n,m)}_{vol}&=&\,\sum_{l=1}^{n+m}\,k^{1-l}(l-1)!\,\sum_{q=0}^l\,S(n,l-q)\,(\lambda+\nu)^{l-q}\,S(m,q)\,(\bar{\lambda}+\bar{\nu})^{q}\, .
\end{eqnarray}
The volume fluctuations correlate those of protons and antiprotons. Thus, e.g., the covariance is non-zero
\begin{equation}
    C_{vol}^{(1,1)}=\frac{(\nu+\lambda)(\bar{\nu}+\bar{\lambda})}{k}\, ,
\end{equation}
in contrast to the one in a system at constant volume, where $C^{(1,1)}_{true}=0$. The net proton variance is then given by (\ref{eq:net-cumulant-2})
\begin{equation}
\kappa_2^{vol}=\nu+\lambda+\bar{\nu}+\bar{\lambda}+\frac{1}{k}\,(\nu+\lambda-\bar{\nu}-\bar{\lambda})^2\, .
\end{equation}

Finally, the CBWC corrected mixed cumulants are given by (see Eq. \eqref{eq:CBWC-mixed-cumulants})
\begin{eqnarray}
    C^{(n,m)}_{CBWC}&=&\sum_{l=1}^{n+m}\bigg[
    \frac{(l-1)!}{(k+\mu+\lambda+\bar{\lambda})^{l-1}}\,\sum_{q=0}^l\,S(n,l-q)\,S(m,q)\,\nu^{l-q}\,\bar{\nu}^{q}\\
    &+&\frac{(-1)^{l-1}(l-1)!}{(\mu+\lambda+\bar{\lambda})^{l-1}}\,\sum_{q=0}^l\,S(n,l-q)\,S(m,q)\,\lambda^{l-q}\,\bar{\lambda}^{q}\bigg]\, .\nonumber
\end{eqnarray} 

We observe that also the CBWC corrected proton-antiproton covariance, 
\begin{align}\label{eq:cbwc_covariance}
    C^{(1,1)}_{CBWC}= \frac{\nu  \bar{\nu} }{k+\lambda +\bar{\lambda} +\mu }-\frac{\lambda  \bar{\lambda} }{\lambda +\bar{\lambda}+\mu }\, ,
\end{align}
is in general non-zero. Furthermore, in a system at constant volume, obtained by taking the limit $k \to \infty$, it is negative, while the true model covariance vanishes, as noted above. This indicates that in certain cases the CBWC procedure may overcompensate the volume fluctuations and thus underestimate the true fluctuations. We return to this issue below.

The first four CBWC corrected net-proton cumulants are then
\begin{eqnarray}\label{eq:CBWC-cumulants-net}
     \kappa_1^{CBWC}&=&\nu-\bar{\nu}+\lambda-\bar{\lambda}\, ,\\
     \kappa_2^{CBWC}&=&\nu+\bar{\nu}+\lambda+\bar{\lambda}+\frac{(\nu-\bar{\nu})^2}{k+\lambda+\bar{\lambda}+\mu}-\frac{(\lambda-\bar{\lambda})^2}{\lambda+\bar{\lambda}+\mu}\, ,\nonumber\\
     \kappa_3^{CBWC}&=&\nu-\bar{\nu}+\lambda-\bar{\lambda}+3\,\bigg(\frac{\nu^2-\bar{\nu}^2}{k+\lambda+\bar{\lambda}+\mu}-\frac{\lambda^2-\bar{\lambda}^2}{\lambda+\bar{\lambda}+\mu}\bigg)\nonumber\\
     &+&2\,\bigg(\frac{(\nu-\bar{\nu})^3}{(k+\lambda+\bar{\lambda}+\mu)^2}+\frac{(\lambda-\bar{\lambda})^3}{(\lambda+\bar{\lambda}+\mu)^2}\bigg)\, ,\nonumber\\
     \kappa_4^{CBWC}&=&\nu+\bar{\nu}+\lambda+\bar{\lambda}
     + \frac{7\, (\nu+\bar{\nu})^2-16\, \nu\,\bar{\nu}}{k+\lambda+\bar{\lambda}+\mu}-\frac{7\, (\lambda+\bar{\lambda})^2-16\, \lambda\,\bar{\lambda}}{\lambda+\bar{\lambda}+\mu}\nonumber\\
     &+&12\,\left(\frac{(\nu-\bar{\nu})^2(\nu+\bar{\nu)}}{(k+\lambda+\bar{\lambda}+\mu)^2}+\frac{(\lambda-\bar{\lambda})^2(\lambda+\bar{\lambda)}}{(\lambda+\bar{\lambda}+\mu)^2}\right)+6\left(\frac{(\nu-\bar{\nu})^4}{(k+\lambda+\bar{\lambda}+\mu)^3}-\frac{(\lambda-\bar{\lambda})^4}{(\lambda+\bar{\lambda}+\mu)^3}\right)\, .\nonumber
 \end{eqnarray}

\section{Discussion}\label{sect:Discussion}
We have examined the efficacy of the CBWC procedure for suppressing ordinary volume fluctuations in a simple model, which allows for an analytical treatment. While the model addresses the effect of correlations between the (net) proton number and multiplicity due to the decay of baryon resonances, it lacks critical dynamics and other potentially pertinent interactions. It is therefore at best semi-realistic and should not be directly confronted with experimental data, but rather be viewed as a convenient tool for exploring potential limitations of the CBWC method.

Let us start the discussion by a simple and extreme example. Suppose that the number of particles
$N$ and the multiplicity $M$ fluctuate, but
$N$ is always proportional to $M$,
\begin{align}
N = c \, M;  \;\;\; c=\rm constant.
  \label{}
\end{align}
By assumption, the true cumulants of $N$ are non-zero.
If we again use multiplicity bins of unit width, the multiplicity in a given bin $j$ is always
$M_{j}$, so that in each bin the multiplicity and consequently also the particle number do not fluctuate. As a result, all
higher order cumulants in each bin vanish and so do the CBWC corrected cumulants for $n\geq 2$, 
\begin{align}
  \kappa_{n\geq 2}[M_{j}] &= 0,\, , \\
  \kappa_{n\geq 2}^{\cb} &=\sum_{j}p_{j} \kappa_{n}[M_{j}] = 0\, ,
  \label{}
\end{align}
in contrast to the true cumulants.
For net proton fluctuations, the implications of such a scenario are somewhat different, as we discuss below.

Such extreme dynamical correlations between
particle number and multiplicity would be realized in the far-fetched situation where all protons
and pions arise exclusively from the decay of $\Delta$-resonances, while the number of primordial
pions and protons both vanish. Nevertheless, this example indicates that the CBWC procedure may underestimate the
magnitude of cumulants in cases where the system exhibits a sufficiently strong correlation between
multiplicity and particle number. To explore this, we examine the results for the second
order cumulants, Eq.~\eqref{eq:k2_all}. As discussed above,  in our model the strength of the dynamical correlations
is determined by the parameter $\lambda$ (i.e., effectively by the number of $\Delta$-resonances). For large
values of $\lambda$ one finds
\begin{align}
  \kappa_2^{CBWC}&=\lambda+\nu+\frac{\nu^2}{k+\lambda+\mu}-\frac{\lambda^2}{\lambda+\mu}
                   \underset{\lambda \gg \nu }{\longrightarrow}  \lambda+\nu
                   -\frac{\lambda^2}{\lambda+\mu} < \kappa_{2}^{true} = \lambda+\nu\, ,
  \label{}
\end{align}
with corresponding results for $\kappa_{3}$ and $\kappa_{4}$ and higher cumulants.
Thus, for large $\lambda$ the CBWC-procedure underestimates the ``true'' cumulants  for a system at fixed volume, $\kappa_n^{true}$. Indeed for $\nu=\mu=0$, i.e., in absence of any primordial protons and pions, we recover the
results of the extreme scenario, where the CBWC corrected cumulants vanish for $n\geq 2$.

\begin{figure}[ht]
    \centering
     \includegraphics[width=0.45\textwidth]{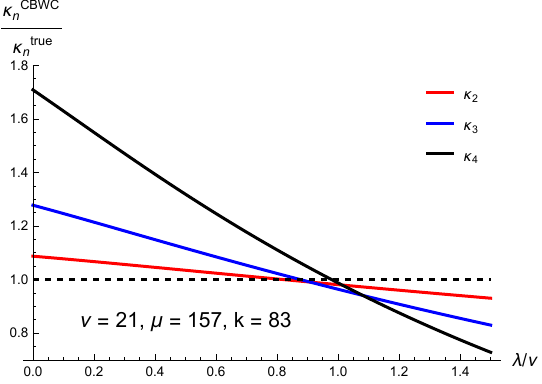}
     \includegraphics[width=0.45\textwidth]{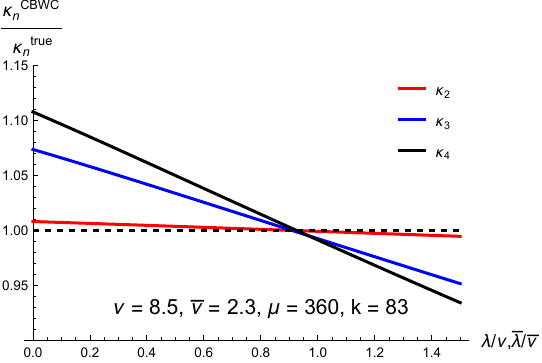}\\
     \includegraphics[width=0.45\textwidth]{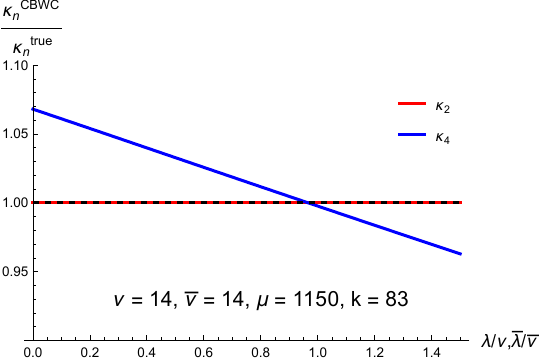}
    \caption{Ratio of CBWC cumulants to their true values, $\kappa_n^{CBWC}/\kappa_n^{true}$, as a function of the correlation strength. Top left: For protons (corresponding to 7.7 GeV collisions). Top right: For net-protons at $\mu_B>0$ (corresponding to 39 GeV collisions). Bottom: For net-protons at $\mu_B=0$ (corresponding to LHC energy collisions).
    }
    \label{fig:kn_vs_lambda}
\end{figure}

\begin{figure}[ht]
    \centering
     \includegraphics[width=0.45\textwidth]{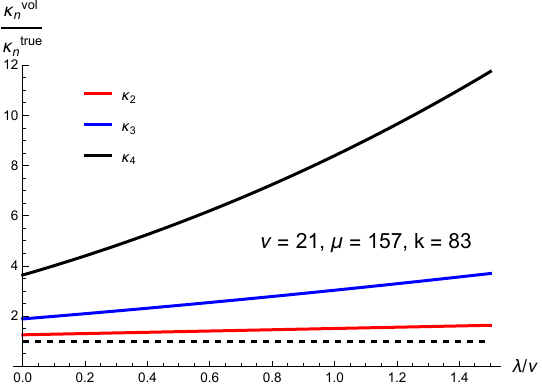}
     \includegraphics[width=0.45\textwidth]{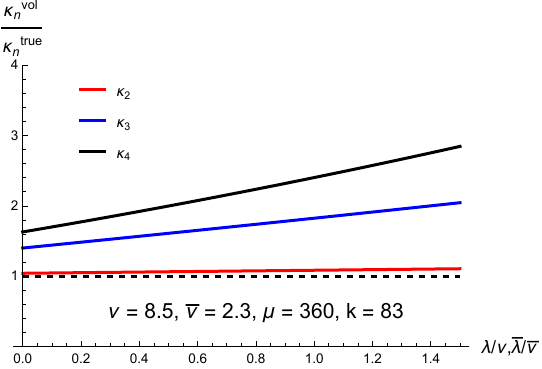}\\
     \includegraphics[width=0.45\textwidth]{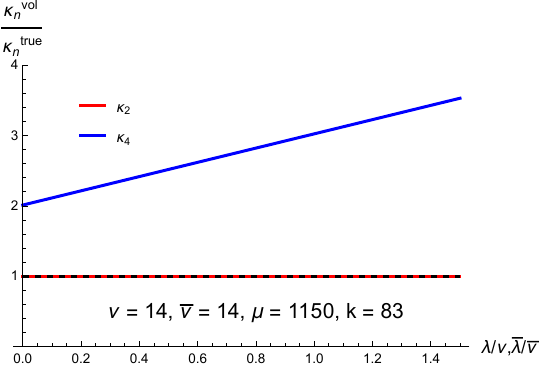}
    \caption{Ratio of cumulants with volume fluctuations over their true value, $\kappa_n^{vol}/\kappa_n^{true}$, as a function of the correlation strength. Top left: For protons (corresponding to 7.7 GeV collisions). Top right: For net-protons at $\mu_B>0$ (corresponding to 39 GeV collisions) . Bottom: For net-protons at $\mu_B=0$ (corresponding to LHC energy collisions).
    }
    \label{fig:kn_vs_lambda_vol}
\end{figure}

The specific value of $\lambda$, above which the
magnitude of $\kappa_{n}^{true}$ is underestimated, varies slowly with the order of the cumulant and is
always close to $\lambda  \simeq \nu$. In the physical picture, this takes place when the
number of baryon resonances decaying in protons and charged pions is approximately the same as the number of primordial protons.
In Fig.~\ref{fig:kn_vs_lambda} we show the ratio of CBWC corrected cumulants over their true values as functions of the correlation strength $\lambda$. Here we have chosen scenarios that in the thermal model of Ref. \cite{Vovchenko:2019pjl} correspond to Au+Au collisions at 7.7 and 39 GeV with the values for temperature and chemical potential, $T=139 \mev,\;\; \mu_B=420 \mev$ and $T=165 \mev,\;\; \mu_B=112 \mev$, respectively. We also show results for LHC energies, corresponding to $T=155 \,\rm{MeV},\;\; \mu_B=0 \mev$. The corresponding mean values for thermal (anti) protons, pions and resonances are given in the figures. A comparison with the ratios of the cumulants with volume fluctuation over their true values, $\kappa^{vol}_{n}/\kappa^{true}_{n}$, shown in Fig.~\ref{fig:kn_vs_lambda_vol}, indicates that the volume fluctuations are indeed substantially reduced by the CBWC procedure. However, for strong correlations the method tends to overcompensate. Moreover, the remaining contamination of $5-10 \%$  at moderate correlation strengths may seem small, but is of the same order of magnitude as the deviations from non-critical baselines observed in the high statistics data of the STAR collaboration \cite{STAR:2025zdq}.

\begin{figure}[ht]
    \centering
     \includegraphics[width=0.45\textwidth]{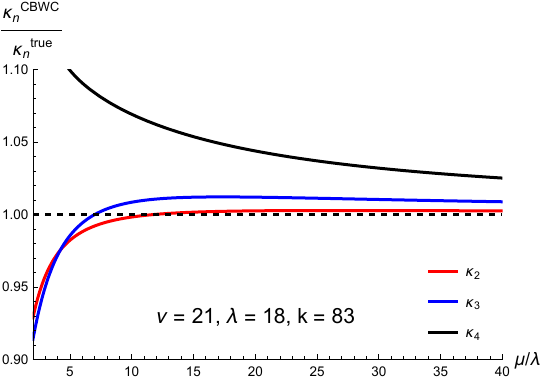}
     \includegraphics[width=0.45\textwidth]{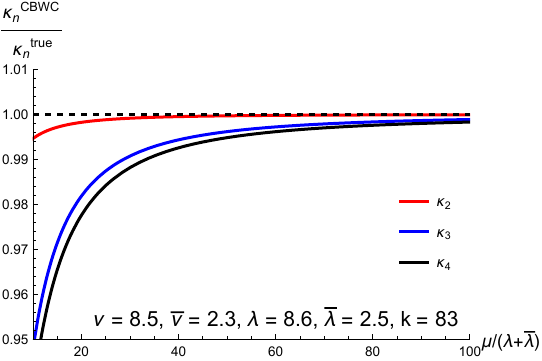}
    \caption{Ratio of CBWC cumulants to their true values, $\kappa_n^{CBWC}/\kappa_n^{true}$, as functions of charged particle multiplicity, $\mu$. Left: For protons. Right: For net protons at $\mu_B>0$. }
    \label{fig:kn_vs_mu}
\end{figure}

In the absence of any dynamical correlations between $N$ and $M$, i.e., when $\lambda = 0$ (in our example, no
$\Delta$-resonances), we find that for large multiplicity, the CBWC variance approaches the desired
result,
\begin{align}
  \kappa_2^{CBWC}&=\nu+\frac{\nu^2}{k+\mu}  \underset{\mu \rightarrow \infty }{\longrightarrow} \kappa_{2}^{true}.
  \label{}
\end{align}
Consequently, the CBWC procedure indeed removes the effect of volume fluctuations in this case.
This is also the case in the limit of $\mu \gg \lambda,\nu$, i.e., when we have many more primordial pions than baryon resonances (decaying into protons and charged pions) and primordial protons. This is demonstrated in Fig.~\ref{fig:kn_vs_mu} where we plot the ratio of CBWC cumulants over the true ones as a function of the mean charged particle multiplicity, $\mu$.

Furthermore, it is worthwhile to explore the effect of the CBWC corrections in a system at fixed volume, i.e., in the absence of volume fluctuations. This case is obtained 
by taking the limit $k\rightarrow \infty$ of the CBWC cumulants $\kappa_{n}^{\cb}$ or factorial cumulants $F_n^{\cb}$ in
Eqs.~\eqref{eq:cumulants_CBWC} and \eqref{eq:CBWC-fact-cum}, respectively. For instance, both the  CBWC variance 
and second-order factorial cumulant are then less than the corresponding true model values
\begin{align}
    \kappa_2^{CBWC}&=\lambda+\nu -\frac{\lambda^2}{(\lambda+\mu)} \leq \kappa_2^{true} = \lambda+\nu\, ,\\
    F_2^{\cb}&=-\frac{\lambda^2}{\lambda+\mu}\leq F_2^{true}=0\, .
\end{align}
Consequently, in the extreme scenarios, i) $\lambda\to\infty$ ($N$ and $M$ strongly correlated) and ii) $k\to\infty$, $\lambda>0$ (no volume fluctuations with dynamical correlations), the CBWC-corrected cumulants clearly overcompensate the volume fluctuations. 

The situation is rather similar for net protons, however, with some minor but interesting differences. Let us first consider the extreme case where all protons, antiprotons, and charged particles arise from the decay of deltas and antideltas, i.e., with $\nu=\bar{\nu}=\mu=0$. As before, this scenario corresponds to the limit of very strong correlations, $\lambda,\, \bar{\lambda}\rightarrow \infty$.  In this case 
\begin{eqnarray}\label{CBWC-cumulants-net-large-lam}
    \kappa_1^{CBWC}&\underset{\lambda,\,\bar{\lambda} \rightarrow \infty }{\longrightarrow}&\lambda-\bar{\lambda}\, ,\\
    \kappa_2^{CBWC}&\underset{\lambda,\,\bar{\lambda} \rightarrow \infty }{\longrightarrow}&\frac{4\,\lambda\,\bar{\lambda}}{\lambda+\bar{\lambda}}\, ,\nonumber\\
    \kappa_3^{CBWC}&\underset{\lambda,\,\bar{\lambda} \rightarrow \infty }{\longrightarrow}&-\frac{8\,\lambda\,\bar{\lambda}\,(\lambda-\bar{\lambda})}{(\lambda+\bar{\lambda})^2}\, ,\nonumber\\
    \kappa_4^{CBWC}&\underset{\lambda,\,\bar{\lambda} \rightarrow \infty }{\longrightarrow}&\frac{16\,\lambda\,\bar{\lambda}\,(\lambda^2-4\,\lambda\,\bar{\lambda}+\bar{\lambda}^2)}{(\lambda+\bar{\lambda})^3}\, .
\end{eqnarray} 
while the true cumulants are $\kappa^{true}_n=\lambda+(-1)^{n}\bar{\lambda}$.

Contrary to the case of protons only, here the CBWC-corrected cumulants do not vanish in the limit of strong correlations. This is because, for a given multiplicity, the sum of proton- and antiproton-numbers is fixed in this limit, while their difference is not constrained and thus fluctuates. Actually, fixing the sum results in protons and anti-protons being anti-correlated. As a consequence, the covariance, Eq.\eqref{eq:cbwc_covariance}, is negative in this limit
\begin{align}
    C^{(1,1)}_{CBWC}\underset{\lambda,\,\bar{\lambda} \rightarrow \infty }{\longrightarrow}-\frac{\lambda  \bar{\lambda} }{\lambda +\bar{\lambda} }\, .
\end{align}

Furthermore, it is worth noting that in this limit, (a)  $\kappa_2^{CBWC}<\kappa_2^{true}$, except for $\lambda=\bar{\lambda}$, since $\kappa_2^{CBWC}-\kappa_2^{true}=-(\lambda-\bar{\lambda})^2/(\lambda+\bar{\lambda})$ and (b) that for 
$\lambda>\bar{\lambda}$, the third CBWC cumulant is negative, $\kappa_3^{CBWC}<0$, while the true one is positive, $\kappa_3^{true}>0$ . 
Finally, the fourth-order CBWC corrected cumulant may turn negative in this limit, while the true cumulant would be $\lambda+\bar{\lambda}>0$. This would be the case for 
\begin{equation}
    \frac{1}{\gamma}<\frac{\bar{\lambda}}{\lambda}<\gamma\, ,
\end{equation} 
with $\gamma=2+\sqrt{3}\approx 3.732$.

However, in scenarios where we have more realistic correlations as well as primordial protons (and antiprotons), the CBWC method does not provide a quantitative measure of how reliable the elimination of volume fluctuations is and how to relate the CBWC-corrected
cumulants to the ones one would like to extract, namely those at constant volume, $V_{0}$. Consequently, it would be desirable to have a measure for the quality of the corrections the CBWC procedure generates, such as, e.g., in the method proposed in \cite{Rustamov:2022sqm,Holzmann:2024wyd}.
Finally, we note that the results presented here are not distorted by the fact that, to facilitate an analytical treatment, we employed bins of unit width in our study of the CBWC method. By numerically sampling the multiplicity distributions, we found that a bin width of $\Delta M \simeq 10$  is sufficient to converge to the analytical results obtained here. 

\section{Conclusion}

In summary, we have studied the CBWC method for removing the effect of volume fluctuations on proton and net-proton cumulants in an analytically tractable model. Our model has one crucial feature, the dynamical correlations that may be attributed to the decay of baryon resonances. However, it lacks any additional dynamics, such as those required for the existence of a critical point. Therefore, our model should not be applied for a quantitative comparison with experimental data. Its purpose is rather to reveal possible limitations of the CBWC method. \
It is clear that the CBWC method preserves the raw moments. The cumulants, on the other hand, are different, which of course is necessary in order to suppress the effect of volume fluctuations.

We find that the CBWC correction does reduce volume fluctuations in principle. However, the remaining contamination from volume fluctuations may still be as large as expected signals from a possible critical point. Furthermore, we also see that for sufficiently strong correlations between charged particle multiplicity and (net) proton number, the CBWC method tends to over correct, consequently suppressing the physics of interest. Note that overcompensation starts already at moderate correlation strengths (cf. Fig.~\ref{fig:kn_vs_lambda}). In the opposite (unphysical) case where there are no correlations between multiplicity and proton number, the CBWC-corrected cumulants do converge to the desired results. 
The upshot is, that for realistic scenarios it is difficult to assess the quality of the CBWC correction, and we were not able to identify a measure which would do so. Therefore, we consider it an important challenge to find such a measure, which, as indicated by this study, is crucial for obtaining a quantitative understanding of the observed cumulants. 

Finally, we have not considered effects of global baryon number conservation, which are known to be important \cite{Bzdak:2012an, Vovchenko:2020tsr,Vovchenko:2020gne,Braun-Munzinger:2020jbk}. It would be interesting to see to which extend these modify our conclusions. However, this is beyond the scope of the present work and we leave this for future studies.

\section*{Acknowledgments}
\hspace*{\parindent}
This material is based upon work supported by the U.S. Department of Energy, 
Office of Science, Office of Nuclear Physics, under contract number 
DE-AC02-05CH11231. This work was started at the ExtreMe Matter Institute (EMMI) Rapid Reaction Task Force (RRTF) on ``Fluctuations and Correlations of conserved Charges''. We thank the participants for many helpful discussions and EMMI for the support, which enabled this stimulating meeting. We also acknowledge useful comments by A. Rustamov.

\appendix
\section{Scaled Gamma distribution}
\label{sec:gamma_dist}
The probability density function (PDF) of the Gamma distribution is given by
\begin{equation}
P_{\Gamma}(z)=\frac{\theta^{-k}z^{k-1}e^{-\frac{z}{\theta}}}\, ,{\Gamma(k)}\label{eq:gamma_dist}
\end{equation}
where $k$ is commonly referred to as the shape parameter and $\theta$ is the scale parameter. 
Its cumulants are given by
\begin{equation}
\kappa_{n}^{\Gamma}=k(n-1)!\theta^{n}\label{eq:gamma_cum}\, ,
\end{equation}
 so that the mean and variance are
\begin{align}
\ave z & =k\theta\label{eq:gamma_mean}\, ,\\
\ave{(\delta z)^{2}} & =k\theta^{2}\, ,
\end{align}
and its moments are  
\begin{equation}
\mu_{n}=\theta^{n}\frac{\Gamma(k+n)}{\Gamma(k)}\, .\label{eq:gamma_mom}
\end{equation}

Let us define a ``scaled'' Gamma distribution in terms
of the reduced volume $x=V/V_{0}$, where $V_{0}$ is the average volume. Setting $z=V$
in Eq. (\ref{eq:gamma_dist}) and $V_{0}=\ave z=k\theta$ so that
$\theta=V_{0}/k$ we get for the PDF of the reduced volume distribution:
\begin{equation}
Q_V(V)=\frac{1}{V_{0}}\frac{k^{k}}{\Gamma(k)}\left(\frac{V}{V_{0}}\right)^{k-1}e^{-\frac{V}{V_{0}}k}=\frac{1}{V_{0}}\frac{k^{k}}{\Gamma(k)}x^{k-1}e^{-xk}.
\end{equation}
We note that the dimension of $Q_V(V)\sim1/{\rm volume}$ so that $\int dV\,Q_V(V)$
is dimensionless as it should. 
Integration over the volume gives
\begin{equation}
\int dV\,Q_V(V)=V_{0}\int dx\,Q_V(V)=\int dx\,\frac{k^{k}}{\Gamma(k)}x^{k-1}e^{-xk}\equiv\int dx\,Q(x)\, ,
\end{equation}
where the Gamma distribution for the reduced volume,
$Q(x)$, is given by
\begin{align}
Q(x) & =\frac{k^{k}}{\Gamma(k)}x^{k-1}e^{-xk}\, ,\non
\int_{0}^{\infty}dx\,Q(x) & =1\, .
\end{align}
Note, that $Q(x)$ is simply a Gamma distribution, Eq. \ref{eq:gamma_dist}
with $z=x$ and $\theta=1/k$. Thus the cumulants of $Q(x)$ are (cf.
Eq. \ref{eq:gamma_cum})
\begin{align}\label{eq:vol-kappa}
\kappa_{n}^{Q} & =\frac{(n-1)!}{k^{n-1}}\, .
\end{align}

\section{Bi-variate Poisson distribution}
\label{sec:bi-variate-poiss}
Two random numbers $N$ and $M$ following a bi-variate Poisson distribution are defined as $N=n+l$  and
$M=m+l$ where $(n,m,l)$  are distributed according to three independent Poisson distributions with
means $(\nu,\mu,\lambda)$, respectively. Therefore the bi-variate probability density function (PDF) can be written as \cite{Campbell:1934,Kawamura:1973ab}
\begin{align}
P_{BP}\left[N, M;\nu,\mu,\lambda\right] &= \sum_{n,m,l=0}^{\infty}P_{P}(n;\nu) P_{P}(m;\mu)
                                     P_{P}(l;\lambda)\, \delta_{N,n+l} \,\delta_{M,m+l} \non
  &= \,e^{-\nu+\mu+\lambda}  \frac{\nu^{N}}{N!}  \frac{\mu^{M}}{M!} \sum_{l=0}^{{\rm Min}(M,N)} l!\,
    {N\choose l} {M \choose l} \left( \frac{\lambda}{\mu \, \nu} \right)^{l} \, ,
\label{eq:bi-poisson_pdf}
\end{align}
where $P_{P}(k;X)=e^{-X}\frac{X^{k}}{k!}$ is the Poisson distribution of $k$ with a mean of
$\ave{k}=X$. The upper limit for the sum over $l$ equal to ${\rm Min}(N,M)$ follows from the restriction of the sums over $n$ and $m$ in the first line to values $\geq 0$.

The moment generating function $h_{BP}(t,s)$ is
\begin{align}
  h_{BP}(t,s) &= \sum_{N,M}P\left[N, M;\nu,\mu,\lambda\right] e^{t \, N}\, e^{s\,M} \non
            &= h_{P}(t;\nu)\, h_{P}(s;\mu)\, h_{P}(t+s;\lambda) \non
  \label{eq:mom_gen_derive}           
\end{align}
and the corresponding cumulant generating function $g_{BP}(t,s)=\log\left[h_{BP}(t,s)\right]$. Here
\begin{align}
h_{P}(t;\nu) = \sum_{n}P_{P}(n;\nu)\,e^{t\, n} = \exp\left[ \left( e^{t}-1 \right) \nu \right]
  \label{}
\end{align}
is the moment generating function for a Poisson distribution with mean $\nu$. The cumulants of the bi-variate Poisson distribution are 
\begin{equation}
    C_{BP}^{(n,m)}= \lambda\,(1-\delta_{n,0}\,\delta_{m,0})+\nu\,\delta_{m,0}\,(1-\delta_{n,0})+\mu\,\delta_{n,0}\,(1-\delta_{m,0}).
\end{equation}

\section{Calculation of CBWC corrected cumulants}
\label{sec:cum_calc}
Here we present some of the details on determining the CBWC corrected cumulants in our model.
As discussed in section \ref{sec:definitions}, in order to compute the CBWC corrected cumulants we need the cumulants for each bin. We consider only
bins of unit width, i.e., bins with a fixed multiplicity $M=M_j$, as this enables us to carry out the necessary sum over bins~\footnote{As noted in Sect.~\ref{sect:Discussion}, this choice does not limit the practical applicability of our results}. To this end,
we define an auxiliary function
\begin{align}
  k(t,M_{j}) &= \sum_{N} P(N, M_{j})\, e^{N\,t} \non
                  & = \int_{-\pi}^{\pi}  \frac{d u}{2 \pi} \, \sum_{N,M} P(N, M)\, e^{N\,t}\, e^{i\,u (M-M_{j})}\non
                  &  =  \int_{-\pi}^{\pi}  \frac{d u}{2 \pi}\,  e^{-i\,u M_{j}} \int_0^\infty dx\, Q(x)\,
                    H_{N,M}(t,i\,u,x)  \, .
\label{eq:h_tilde}
\end{align}
Here we used the representation of the Kronecker delta, $\delta_{m,n}=\int_{-\pi}^{\pi}\,  \frac{d u}{2 \pi}\,e^{i\,u (m-n)}$ and the definition of the volume-dependent moment generating function,
Eq.~\eqref{eq:mom_gen_all}. We also need the probability $p_{i}$ to be in bin $i$, which in our case
of bins of unit width is simply
\begin{align}
p_{i}=\sum_{N} P(N, M_{i}) = k(t=0,M_{i})\, .
  \label{eq:bin_prob_model}
\end{align}

Using Eqs. \eqref{eq:part_dist_bin} and \eqref{eq:mom_gen_bin}, the generating function of moments in bin $j$
is 
\begin{align}
h(t,M_{j}) = \frac{1}{p_{j}} \sum_{N}P(N, M_{j})\, e^{N\,t}= \frac{k(t,M_{j})}{k(0,M_{j})}\, .
  \label{eq:mom_gen_from_h_tilde}
\end{align}
To evaluate $k(t,M_{j})$, we first carry out the Fourier transform
\begin{align}
  F(M_{j},x,t) &=  \int_{-\pi}^{\pi}  \frac{d u}{2 \pi}\, e^{-i\,u\,M_{j}}\,H_{N,M}(t,i\,u,x) \non
           &= \frac{\left((e^t\,\lambda+\mu)\,x\right)^{M_{j}}}{M_{j}!}\,e^{-\left(\lambda+\mu+(1-e^t)\nu\right)\,x}\, .
  \label{eq:fourier_transform_h}
\end{align}
Integrating out the volume fluctuations then yields
\begin{align}
  k[t,M_{j}]   &=  \int dx\, Q(x)\, F(M_{j},x,t) \non
                       &=  \frac{\Gamma(M_{j}+k)}{\Gamma(k)\,\Gamma(M_{j}+1)}\,
                               \frac{k^k\,(e^t\,\lambda+\mu)^{M_{j}}}{(k+\lambda+\mu+(1-e^t)\nu)^{M_{j}+k}}\, .
  \label{eq:h_tilde_final}
\end{align}

The moment generating function for a given bin is then obtained from  Eq.\eqref{eq:mom_gen_from_h_tilde},
\begin{align}
  h(t,M_{j}) = 
           \left(\frac{e^t\,\lambda+\mu}{\lambda+\mu}\right)^{M_{j}}\,\left(\frac{k+\lambda+\mu}{k+\lambda+\mu+(1-e^t)\nu}\right)^{M_{j}+k}\, ,
  \label{}
\end{align}
with the corresponding cumulant generating function
\begin{align}
    g(t;M_{j})=M_{j}\,\log\left(\frac{e^t\,\lambda+\mu}{\lambda+\mu}\right)-(M_{j}+k)\,\log\left(\frac{k+\lambda+\mu+(1-e^t)\nu}{k+\lambda+\mu}\right)\, .
  \label{eq:cumulant-gen-function-app}
\end{align}

\section{Net-protons}
\label{sec:appendix_net_prot}

In order to include the fluctuations of antiprotons in the analysis, we extend the model presented in section \ref{sec:Analysis-CBWC} and appendices \ref{sec:gamma_dist} - \ref{sec:cum_calc}. The strategy we follow here is to determine the mixed cumulants, which are obtained by differentiating the mixed cumulant generating function $G(t,s)$, (cf. Eq.~\eqref{eq:mixed-gen-func})
\begin{align} \label{eq:mixed_cum_net}
    C^{(n,m)} = \frac{\partial^n}{\partial t^n}\,\frac{\partial^m}{\partial s^m}\,G(t,s)|_{t=0,s=0}\, .   
\end{align}
The net cumulants are then obtained using (\ref{eq:net-cumulant}).

In order to construct the probability distribution for the extended model, we start with a product of five Poisson distributions
\begin{equation}
P_{5P}\left[n,\bar{n},m,l,\bar{l}\right]=P_P(n;\nu)\,P_P(\bar{n};\bar{\nu})\,P_P(m;\mu)\,P_P(l,\lambda)\,P_P(\bar{l};\bar{\lambda})\, ,
\end{equation}
using the notation introduced in appendix \ref{sec:bi-variate-poiss}.
We then define a tri-variate probability distribution for the number of protons $N$, antiprotons $\bar{N}$ and multiplicity $M$
\begin{align}
    P_{TP}\left[N,\bar{N},M;\nu,\bar{\nu},\mu,\lambda,\bar{\lambda}\right]&=\sum_{l=0}^{{\rm Min}(N,M)}\sum_{\bar{l}=0}^{{\rm Min}(\bar{N},M-l)}\,P_{5P}\left[N-l,\bar{N}-\bar{l},M-l-\bar{l},l,\bar{l}\right]\\ 
    &=e^{-(\nu+\bar{\nu}+\mu+\lambda+\bar{\lambda})}\,\sum_{l=0}^{{\rm Min}(N,M)}\,\sum_{\bar{l}=0}^{{\rm Min}(\bar{N},M-l)} \frac{\nu^{\,N-l}\,}{l!\,(N-l)!}
    \frac{\bar{\nu}^{\,\bar{N}-\bar{l}}}{\bar{l}!\,(\bar{N}-\bar{l})!}\,\frac{\mu^{\,M-l-\bar{l}}\,\lambda^{\,l}\,\bar{\lambda}^{\,\bar{l}}}{(M-l-\bar{l})!}\, . \nonumber 
\end{align}
The parameter $\lambda$ governs the strength of the correlation between the proton number and the multiplicity, while $\bar{\lambda}$ is responsible for the correlation between antiprotons and multiplicity.
The moment generating function for this distribution is given by
\begin{align}\label{eq:tri-poisson-mom-gen-func}
    h_{TP}(t,s,u)&=\sum_{N,\bar{N},M}\,e^{N\,t+\bar{N}\,s+M\,u}\,P_{TP}\left[N,\bar{N},M;\nu,\bar{\nu},\mu,\lambda,\bar{\lambda}\right]\\
&=h_P(t;\nu)\,h_P(s;\bar{\nu})\,h_P(u;\mu)\,h_P(t+u;\lambda)\,h_P(s+u;\bar{\lambda})\, .\nonumber
\end{align} 

The moment generating function for the true fluctuations is obtained by setting $u=0$ in \eqref{eq:tri-poisson-mom-gen-func}. The corresponding cumulant generating function is then
\begin{align} 
\label{eq:cum_gen_net_true}
G_{true}(t,s) = \log\left[h_{TP}(t,s,0)\right]
= (e^t-1)\,(\lambda+\nu)+(e^s-1)\, (\bar{\lambda}+\bar{\nu}) \, ,
\end{align}
which corresponds to that of a product of two independent Poissonians with means $\lambda+\nu$ and $\bar{\lambda}+\bar{\nu}$. 
The mixed cumulants are then obtained using \eqref{eq:mixed_cum_net}
\begin{align} \label{eq:mixed_cum_net_true}
    C^{(n,m)}_{true} =(\nu+\lambda)\delta_{m,0}(1-\delta_{n,0})+(\bar{\nu}+\bar{\lambda}
)\delta_{n,0} (1-\delta_{m,0})\, , 
\end{align}
while the corresponding net cumulants are those of the Skellam distribution~\cite{Skellam:1946} with means $\nu+\lambda$ and $\bar{\nu}+\bar{\lambda}$,
\begin{equation}
    \kappa^{true}_n=\nu+\lambda + (-1)^n\,(\bar{\nu}+\bar{\lambda})\, .
\end{equation}

As in Section \ref{sec:Analysis-CBWC}, we assume that the volume fluctuations are described by a gamma distribution of the ratio of the volume $V$ to its average, $x=V/V_0$,
\begin{equation}\label{eq:wound-nucl-gamma-2}
    Q(x)=\frac{k^k}{\Gamma(k)}\,x^{(k-1)}e^{-k\,x}\, .
\end{equation}
The probability distribution in a given volume $V=x\,V_0$ is again obtained by scaling the parameters $\nu,\,\bar{\nu},\,\mu,\,\lambda$ and $\bar{\lambda}$ by the reduced volume $x$,
\begin{equation}
P_{TP}\left[N,\bar{N},M;x\right]\equiv P_{TP}\left[N,\bar{N},M;x\,\nu,x\,\bar{\nu},x\,\mu,x\,\lambda,x\,\bar{\lambda}\right]\, .
\end{equation}
The corresponding moment generating function is then given by
\begin{align}
    h_{TP}(t,s,u;x)
    &=h_P(t;x\,\nu)\,h_P(s;x\,\bar{\nu})\,h_P(u;x\,\mu)\,h_P(t+u;x\,\lambda)\,h_P(s+u;x\,\bar{\lambda})\, ,
\end{align}
with $h_P(t,X)=\exp[X (e^{t}-1)]$ being the moment generating function of a Poissonian with mean $X$.
Moreover, the generating function for moments of the proton and antiproton distributions and the Fourier transform of the charged particle distribution is
\begin{align}
    \hat{h}_{TP}(t,s,u;x)&=\sum_{N,\bar{N},M}\,e^{N\,t+\bar{N}\,s+i\,M\,u}\,P_{TP}\left[N,\bar{N},M;x\right]\\
    &=e^{-(\nu+\bar{\nu}+\mu+\lambda+\bar{\lambda})\,x}e^{(e^{i\,u}(\mu+e^t\,\lambda+e^s\,\bar{\lambda})+e^t\,\nu+e^s\,\bar{\nu})\,x}\, .
\end{align} 

We are now ready to write down the moment generating function for protons and antiprotons at fixed multiplicity
\begin{align}\label{eq:mom-gen-func-fixed-M}
    h_{TP}(t,s;M)=\frac{k(t,s;M)}{k(0,0;M)}\, ,
\end{align}
where
\begin{align}\label{eq:moment-gen-fixed-M-2}
    k(t,s;M)&=\int\,dx\,Q(x)\,\sum_{N,\bar{N}}P_{TP}(N,\bar{N},M;x)\exp(N \,t + \bar{N}\,s )& \\ \nonumber
    &=\int\,dx\,Q(x)\,F_{TP}(M,x,t,s)\, ,
\end{align}
and
\begin{align}\label{eq:fourier-trans-h-hat-2}
    F_{TP}(M,x,t,s)&=\int_{-\pi}^\pi\frac{d\,u}{2\,\pi}\,\exp\left[-i\,u\,M\right]\,\hat{h}_{TP}(t,s,u,x)\\
    &=\frac{((\mu+e^t\,\lambda+e^s\,\bar{\lambda})x)^M}{M!}\,e^{((e^t-1)\nu+(e^s-1)\bar{\nu}-\mu-\lambda-\bar{\lambda})\,x}\, .\nonumber
\end{align}
This implies that
\begin{align}
    k(t,s;M)
    &=\frac{\Gamma(M+k)}{\Gamma(k)\,\Gamma(M+1)}\frac{k^k\,(\mu+e^t\,\lambda+e^s\,\bar{\lambda})^M}{(k+\mu+\lambda+\bar{\lambda}+(1-e^t)\nu+(1-e^s)\bar{\nu})^{M+k}}\, .
\end{align}
Analogous to Eq.\eqref{eq:bin_prob_model} the statistical weight of each bin is given by
\begin{align}
p_i=k(0,0,M_i)\, .
\label{eq:bin_prob_net}
\end{align}
The moment generating function for fixed $M$ is then
\begin{align}
    h_{TP}(t,s;M)&=\left(\frac{\mu+e^t\,\lambda+e^s\,\bar{\lambda}}{\mu+\lambda+\bar{\lambda}}\right)^M \left(\frac{k+\mu+\lambda+\bar{\lambda}}{k+\mu+\lambda+\bar{\lambda}+(1-e^t)\nu+(1-e^s)\bar{\nu}}\right)^{M+k}\, ,
\end{align}        
and the corresponding cumulant generating function 
\begin{align}\label{eq:gen-func-fixed-M}
    g_{TP}(t,s;M)&=M\,\log\left(\frac{\mu+e^t\,\lambda+e^s\,\bar{\lambda}}{\mu+\lambda+\bar{\lambda}}\right)-(M+k)\,\log\left(\frac{k+\mu+\lambda+\bar{\lambda}+(1-e^t)\nu+(1-e^s)\bar{\nu}}{k+\mu+\lambda+\bar{\lambda}}\right)\, .
\end{align}

Summing over multiplicities in (\ref{eq:moment-gen-fixed-M-2}), one obtains the total moment generating function (including volume fluctuations)
\begin{equation}\label{eq:total-gen-function-2}
    H_{vol}(t,s)=\sum_{M=-\infty}^\infty k(t,s;M)=\int\,dx\,Q(x)\,\hat{h}_{TP}(t,s,0,x)\, ,
\end{equation}
where 
\begin{align}\label{eq:sum-Q-2}
    \hat{h}_{TP}(t,s,0,x)&=\sum_{M=-\infty}^\infty F(M,x,t,s)\\
    &=\exp\left([(e^t-1)\,(\lambda+\nu)+(e^s-1)\,(\bar{\lambda}+\bar{\nu})]\,x\right)\, ,\nonumber
\end{align}
is the moment generating function for a product of two independent Poisson distributions with means $(\lambda+\nu)\, x$ and $(\bar{\lambda}+\bar{\nu})\, x$, respectively.

Evaluating the integral over the reduced volume in (\ref{eq:total-gen-function-2}) we find
\begin{align}
    H_{vol}(t,s)
    &=\left(\frac{k}{k+(1-e^t)\,(\lambda+\nu)+(1-e^s)\,(\bar{\lambda}+\bar{\nu})}\right)^k \, ,
\end{align}
while the corresponding generating function for the total cumulants is
\begin{align}\label{eq:cumulant-gen-func}
    G_{vol}(t,s)&=\log(H_{vol}(t,s))=-k\,\log\left(\frac{k+(1-e^t)\,(\lambda+\nu)+(1-e^s)\,(\bar{\lambda}+\bar{\nu})}{k}\right)\, .
\end{align}

The mixed cumulants can be computed by employing the Fa\'a di Bruno formula for the derivatives of a composite function of several variables \cite{Riordan:1946,Friman:2022wuc} $f(g(z_1,z_2))$
\begin{eqnarray}\label{eq:faa-di-bruno-mult1}
    \frac{\partial^{n}}{\partial t^{n}}\,\frac{\partial^{m}}{\partial s^{m}}f(g(t,s))=\sum_{l=1}^{n+m}f^{(l)}\,B_{n,m;l}(\{g^{(i,j)}\})\, ,
\end{eqnarray}
where $f(x)=-k\log(x)$ and $g(t,s)=1+\left[(1-e^t)\,(\lambda+\nu)+(1-e^s)\,(\bar{\lambda}+\bar{\nu})\right]/k$.
The derivatives are 
\begin{equation}\label{eq:f-derivatives}
    f^{(l)}=k\,(-1)^{l}\,(l-1)!    
\end{equation}
and 
\begin{equation}\label{eq:g-derivatives}
    g^{(i,j)}=-\frac{\lambda+\nu}{k}\,\delta_{j,0}-\frac{\bar{\lambda}+\bar{\nu}}{k}\,\delta_{i,0}\, .
\end{equation}
The bivariate Bell polynomials are given by \cite{Schumann:2019xy}
\begin{equation}
    B_{n,m;l}=\frac{1}{l!}\,\frac{\partial^{n}}{\partial t^{n}}\frac{\partial^{n}}{\partial s^{m}}\frac{\partial^{l}}{\partial u^{l}}\,\Phi(t,s,u)|_{t=s=u=0}\, ,
\end{equation}
where
\begin{eqnarray}\label{eq:gen-function-multiv-Bell-21}
    \Phi(t,s,u)&=&\exp\left(u\,\sum_{\substack{j_1,j_2\\ j_1+j_2>0}}^\infty x^{(j_1,j_2)}\frac{t^{j_1}\,s^{j_2}}{j_1!\,j_2!}\right)\\
    &=&\exp \Bigg[u\,\left(-(e^t-1)\left(\frac{\lambda+\nu}{k}\right)-(e^s-1)\left(\frac{\bar{\lambda}+\bar{\nu}}{k}\right)\right)\Bigg]\, ,\nonumber
\end{eqnarray}
and $\Phi(t,s,u)$ is the generating function for the bivariate Bell polynomial. In the second line we inserted the derivatives $x^{(i,j)}=g^{(i,j)}$ and evaluated the sums over $j_1$ and $j_2$. Given the generating function for Stirling numbers of the second kind
\begin{equation}
\Psi(t,u)=\exp\Big[u\,\left(e^t-1\right)\Big]\, ,    
\end{equation}
we find that, for the derivatives (\ref{eq:g-derivatives}), the Bell polynomials are given by
\begin{equation}\label{eq:Bell-total}
    B_{n,m;l}(\{g^{(i,j)}\})=\frac{(-1)^l}{k^l}\sum_{q=0}^l\,S(n,l-q)\left(\lambda+\nu\right)^{l-q}\,S(m,q)\,\left(\bar{\lambda}+\bar{\nu}\right)^q\, ,
\end{equation}
where 
\begin{equation}
    S(n,l)=\frac{1}{l!}\,\frac{\partial^n}{\partial t^n}\,\frac{\partial^l}{\partial u^l}\,\Psi(t,u)|_{t=u=0}
\end{equation}
are Stirling numbers of the second kind.
Using (\ref{eq:mixed_cum_net}), (\ref{eq:faa-di-bruno-mult1}), \ref{eq:f-derivatives}) and (\ref{eq:Bell-total}) one then finds
\begin{eqnarray}\label{eq:net-proton-fluct-vol}
    C^{(n,m)}_{vol}&=&\,\sum_{l=1}^{n+m}\,k^{1-l}(l-1)!\,\sum_{q=0}^l\,S(n,l-q)\,(\lambda+\nu)^{l-q}\,S(m,q)\,(\bar{\lambda}+\bar{\nu})^{q}\, .
\end{eqnarray}

The lowest cumulants
\begin{equation}\label{eq:proton-antiproton-cumulant-1}
    C_{vol}^{(1,0)}=\nu+\lambda,\qquad C_{vol}^{(0,1)}=\bar{\nu}+\bar{\lambda}\, ,
\end{equation}
are equal to the proton and antiproton average multiplicities. The net proton cumulants are obtained using (\ref{eq:net-cumulant}). The first one is given by
\begin{equation}\label{eq:net-cumulant-1}
    \kappa^{vol}_1=\nu-\bar{\nu}+\lambda-\bar{\lambda}\, ,
\end{equation}
while the variance of the net proton number is
\begin{equation}\label{eq:net-cumulant-2}
\kappa_2^{vol}=C_{vol}^{(2,0)}+C_{vol}^{(0,2)}-2\,C_{vol}^{(1,1)}=\nu+\lambda+\bar{\nu}+\bar{\lambda}+\frac{1}{k}\,(\nu+\lambda-\bar{\nu}-\bar{\lambda})^2\, .
\end{equation}
Thus, at high energies, where the net proton number vanishes, the contribution from volume fluctuations to the variance (the last term in (\ref{eq:net-cumulant-2})) vanishes, (see Ref.~\cite{Skokov:2012ds}) and we recover the Skellam result.
Similarly, we find for the third and fourth order cumulants with volume fluctuations
\begin{eqnarray}\label{eq:volume-cumulants-3-4}
     \kappa_3^{vol}&=&\nu-\bar{\nu}+\lambda-\bar{\lambda}+\frac{3}{k}\,\big[(\nu+\lambda)^2-(\bar{\nu}+\bar{\lambda})^2\big]
     +\frac{2}{k^2}\,\big(\nu+\lambda-\bar{\nu}-\bar{\lambda}\big)^3\, ,\\
     \kappa_4^{vol}&=&\nu+\bar{\nu}+\lambda+\bar{\lambda}
     + \frac{1}{k}\,\big[7(\nu+\lambda+\bar{\nu}+\bar{\lambda})^2-16( \nu+\lambda)\,(\bar{\nu}+\bar{\lambda})\big]\nonumber\\
     &+&\frac{12}{k^2}\,\big(\nu+\lambda-\bar{\nu}-\bar{\lambda}\big)^2\big(\nu+\lambda+\bar{\nu}+\bar{\lambda}\big)+\frac{6}{k^3}\big(\nu+\lambda-\bar{\nu}-\bar{\lambda}\big)^4\, .\nonumber
 \end{eqnarray}

We now turn to the cumulants at fixed $M$, 
\begin{equation}
     C^{(n,m)}(M)=\frac{\partial^n}{\partial t^n}\,\frac{\partial^m}{\partial s^m}\,g_{TP}(t,s;M)|_{t=0,s=0}\, .
\end{equation}
We again employ the Fa\'a di Bruno formula (\ref{eq:faa-di-bruno-mult1}) with $f_1(x)=M\, \log(x)$ and $f_2(x)=-(M+k)\, \log(x)$ for the first and second term in (\ref{eq:gen-func-fixed-M}), respectively. The corresponding forms of the inner function are
\begin{equation}
    g_1(t,s)=1+\frac{(e^t-1)\,\lambda+(e^s-1)\,\bar{\lambda}}{\mu+\lambda+\bar{\lambda}}\, ,
\end{equation} 
and 
\begin{equation}
    g_2(t,s)=1+\frac{(1-e^t)\nu+(1-e^s)\bar{\nu}}{k+\mu+\lambda+\bar{\lambda}}\, .
\end{equation}
The derivatives of $g_1$ and $g_2$ are then
\begin{eqnarray}
    g_1^{(i,j)}&=&\frac{\lambda}{\mu+\lambda+\bar{\lambda}}\,\delta_{j,0}+\frac{\bar{\lambda}}{\mu+\lambda+\bar{\lambda}}\,\delta_{i,0}\\
    g_2^{(i,j)}&=&-\frac{\nu}{k+\mu+\lambda+\bar{\lambda}}\,\delta_{j,0}-\frac{\bar{\nu}}{k+\mu+\lambda+\bar{\lambda}}\,\delta_{i,0}\, ,
\end{eqnarray} 
and the resulting cumulants are given by
\begin{eqnarray}
    C^{(n,m)}(M)&=&\sum_{l=1}^{n+m}\bigg[M\,\frac{(-1)^{l-1}(l-1)!}{(\mu+\lambda+\bar{\lambda})^l}\,\sum_{q=0}^l\,S_{n,l-q}\,S_{m,q}\,\lambda^{l-q}\,\bar{\lambda}^{q}\\
    &+&(M+k)\,\frac{(l-1)!}{(k+\mu+\lambda+\bar{\lambda})^l}\,\sum_{q=0}^l\,S_{n,l-q}\,S_{m,q}\,\nu^{l-q}\,\bar{\nu}^{q}\bigg]\, .\nonumber
\end{eqnarray} 
We thus find the two lowest cumulants
\begin{align}
    C^{(1,0)}(M)&=\frac{\nu\,(k+M)}{k+\mu+\lambda+\bar{\lambda}}+\frac{\lambda\,M}{\mu+\lambda+\bar{\lambda}}\, ,\\
    C^{(0,1)}(M)&=\frac{\bar{\nu}\,(k+M)}{k+\mu+\lambda+\bar{\lambda}}+\frac{\bar{\lambda}\,M}{\mu+\lambda+\bar{\lambda}}\, ,
\end{align}
and the first net proton cumulant 
\begin{equation}
    \kappa_1(M)=\frac{(\nu-\bar{\nu})\,(k+M)}{k+\mu+\lambda+\bar{\lambda}}+\frac{(\lambda-\bar{\lambda})\,M}{\mu+\lambda+\bar{\lambda}}\, .
\end{equation}

Summing the fixed-$M$ cumulants over $M$ with the statistical weight, Eq.\eqref{eq:bin_prob_net}, we obtain the CBWC cumulants. Since the cumulants are linear in $M$, the multiplicity is replaced by its average $\langle M\rangle =\mu+\lambda+\bar{\lambda}$. We thus recover (\ref{eq:proton-antiproton-cumulant-1}) and (\ref{eq:net-cumulant-1}). Since the cumulant generating function is also linear in the multiplicity, we obtain the generating function for CBWC cumulants by doing the same replacement in (\ref{eq:gen-func-fixed-M}),
\begin{align}\label{eq:gen-func-CBWC}
    G(t,s)_{CBWC}&=(\mu+\lambda+\bar{\lambda})\,\log\left(\frac{\mu+e^t\,\lambda+e^s\,\bar{\lambda}}{\mu+\lambda+\bar{\lambda}}\right)\\
    &-(k+\mu+\lambda+\bar{\lambda})\,\log\left(\frac{k+\mu+\lambda+\bar{\lambda}+(1-e^t)\nu+(1-e^s)\bar{\nu}}{k+\mu+\lambda+\bar{\lambda}}\right)\, .\nonumber
\end{align}
It follows that the CBWC cumulants are given by
\begin{eqnarray}\label{eq:CBWC-mixed-cumulants}
    C^{(n,m)}_{CBWC}&=&\sum_{l=1}^{n+m}\bigg[
\frac{(l-1)!}{(k+\mu+\lambda+\bar{\lambda})^{l-1}}\,\sum_{q=0}^l\,S(n,l-q)\,S(m,q)\,\nu^{l-q}\,\bar{\nu}^{q}\\
&+&\frac{(-1)^{l-1}(l-1)!}{(\mu+\lambda+\bar{\lambda})^{l-1}}\,\sum_{q=0}^l\,S(n,l-q)\,S(m,q)\,\lambda^{l-q}\,\bar{\lambda}^{q}\bigg]\, .\nonumber
\end{eqnarray}
 
 The net proton CBWC cumulants are obtained by applying (\ref{eq:net-cumulant}) to (\ref{eq:CBWC-mixed-cumulants}). We find for the first four cumulants
 \begin{eqnarray}\label{eq:CBWC-cumulants}
     \kappa_1^{CBWC}&=&\nu-\bar{\nu}+\lambda-\bar{\lambda}\, ,\\
     \kappa_2^{CBWC}&=&\nu+\bar{\nu}+\lambda+\bar{\lambda}+\frac{(\nu-\bar{\nu})^2}{k+\lambda+\bar{\lambda}+\mu}-\frac{(\lambda-\bar{\lambda})^2}{\lambda+\bar{\lambda}+\mu}\, ,\nonumber\\
     \kappa_3^{CBWC}&=&\nu-\bar{\nu}+\lambda-\bar{\lambda}+3\,\bigg(\frac{\nu^2-\bar{\nu}^2}{k+\lambda+\bar{\lambda}+\mu}-\frac{\lambda^2-\bar{\lambda}^2}{\lambda+\bar{\lambda}+\mu}\bigg)\nonumber\\
     &+&2\,\bigg(\frac{(\nu-\bar{\nu})^3}{(k+\lambda+\bar{\lambda}+\mu)^2}+\frac{(\lambda-\bar{\lambda})^3}{(\lambda+\bar{\lambda}+\mu)^2}\bigg)\, ,\nonumber\\
     \kappa_4^{CBWC}&=&\nu+\bar{\nu}+\lambda+\bar{\lambda}
     + \frac{7\, (\nu+\bar{\nu})^2-16\, \nu\,\bar{\nu}}{k+\lambda+\bar{\lambda}+\mu}-\frac{7\, (\lambda+\bar{\lambda})^2-16\, \lambda\,\bar{\lambda}}{\lambda+\bar{\lambda}+\mu}\nonumber\\
     &+&12\,\left(\frac{(\nu-\bar{\nu})^2(\nu+\bar{\nu)}}{(k+\lambda+\bar{\lambda}+\mu)^2}+\frac{(\lambda-\bar{\lambda})^2(\lambda+\bar{\lambda)}}{(\lambda+\bar{\lambda}+\mu)^2}\right)+6\left(\frac{(\nu-\bar{\nu})^4}{(k+\lambda+\bar{\lambda}+\mu)^3}-\frac{(\lambda-\bar{\lambda})^4}{(\lambda+\bar{\lambda}+\mu)^3}\right)\, .\nonumber
 \end{eqnarray}

We note that in the high-energy limit, where $\nu=\bar{\nu}$ and $\lambda=\bar{\lambda}$, the odd net CBWC proton cumulants vanish, as do the corresponding true ones. Moreover, in this limit, the second and third terms in the CBWC variance vanish, and hence one recovers the corresponding true net proton result (\ref{eq:net-cumulant-2}). Thus, the contribution from volume fluctuations to the variance vanishes at high energies also for the CBWC result.  

\section{Factorial Cumulants}\label{sec:factorial-cumulants} 
The connection between cumulants and factorial cumulants is in general given by \cite{Friman:2022wuc}
\begin{equation}
 F_n^{a}=\sum_{j=1}^n\,s(n,j)\,\kappa_j^{a}\, ,\qquad   \kappa_n^{a}=\sum_{j=1}^n\,S(n,j)\,F_j^{a}\, .
\end{equation}
Since we are dealing with Poissonians, the true factorial cumulants vanish for $n\geq2$, $F_n^{true}=0;\;\; n\geq2$.
The factorial cumulants with volume fluctuations are given by
\begin{equation}
    F_n^{vol}=(n-1)!\, k\, \left(\frac{\lambda+\nu}{k}\right)^n\, ,
\end{equation}
which yields
\begin{align}
    F_2^{vol}=&\frac{(\lambda +\nu )^2}{k} \, ,\non
    F_3^{vol}=& \frac{2 (\lambda +\nu )^3}{k^2}\\, ,non
    F_4^{vol}=&\frac{6 (\lambda +\nu )^4}{k^3}\, .
\end{align}
Furthermore, the CBWC corrected ones are
\begin{equation}
    F_n^{CBWC}=(n-1)!\,\left(\frac{\nu^n}{(k+\lambda+\mu)^{n-1}}-\frac{(-\lambda)^n}{(\lambda+\mu)^{n-1}}\right)\, ,
\end{equation}
which for the first three non-trivial factorial cumulants yields
\begin{align}
    F_2^{CBWC}=&\frac{\nu ^2}{k+\lambda +\mu }-\frac{\lambda ^2}{\lambda
   +\mu } \, ,\non
    F_3^{CBWC}=&\frac{2 \nu
   ^3}{(k+\lambda +\mu )^2}+\frac{2 \lambda ^3}{(\lambda +\mu )^2} \, ,\non
   F_4^{CBWC}=&\frac{6 \nu ^4}{(k+\lambda +\mu )^3}-\frac{6 \lambda
   ^4}{(\lambda +\mu )^3}\,.
\end{align}

We note that the factorial cumulants of order $n\geq 2$ with volume fluctuations and the CBWC corrected ones are simply the terms with the highest power in the denominator in Eqs.\eqref{eq:k2_all}, \eqref{eq:k3_all}, and \eqref{eq:k4_all}. 
\bibliography{Bib_cbwc_paper}

\end{document}